\documentclass[aps,prl,twocolumn,showpacs,superscriptaddress]{revtex4}
\usepackage{amsmath}
\usepackage{amssymb}
\usepackage{epsfig}
\usepackage{color}
\usepackage{graphicx,amsmath}
\usepackage{bm}

\begin{document}
\newcommand{\beq}{\begin{equation}}
\newcommand{\eeq}{\end{equation}}

\title{Merging of Landau levels in a strongly-interacting two-dimensional electron system in silicon}
\author{A.~A. Shashkin}
\affiliation{Institute of Solid State Physics, Chernogolovka, Moscow District, 142432, Russia}
\author{V.~T. Dolgopolov}
\affiliation{Institute of Solid State Physics, Chernogolovka, Moscow District, 142432, Russia}
\affiliation{Moscow Institute of Physics and Technology, Dolgoprudny, Moscow District 141700, Russia}
\author{J.~W. Clark}
\affiliation{McDonnell Center for the Space Sciences \& Department of Physics, Washington University, St.~Louis, MO 63130, USA}
\author{V.~R. Shaginyan}
\affiliation{Petersburg Nuclear Physics Institute, NRC Kurchatov Institute, Gatchina, 188300, Russia}
\affiliation{Clark Atlanta University, Atlanta, GA 30314, USA}
\author{M.~V. Zverev}
\affiliation{NRC Kurchatov Institute, Moscow, 123182, Russia}
\affiliation{Moscow Institute of Physics and Technology, Dolgoprudny, Moscow District 141700, Russia}
\author{V.~A. Khodel}
\affiliation{NRC Kurchatov Institute, Moscow, 123182, Russia}
\begin{abstract}
We show that the merging of the spin- and valley-split Landau levels at the chemical potential is an intrinsic property of a strongly-interacting two-dimensional electron system in silicon. Evidence for the level merging is given by available experimental data.
\end{abstract}
\maketitle

\date{\today}

In a non-interacting fermion system with continuous spectrum, the occupation probability for a quantum state at fixed chemical potential and temperature is a function of the single-particle energy only \cite{landau}. If the temperature tends to zero, the energy interval separating the filled and empty quantum states also tends to zero. For free particles there appears a Fermi surface in momentum space with dimensionality $d-1$, where $d$ is the dimensionality of the fermions.

In general, this reasoning is not true for interacting fermions \cite{ks,vol,noz,prb2008,shagrep,an2012,zverev12}. In this case the single-particle energy depends on electron distributions, and the occupation numbers of quantum states at the chemical potential can be different, falling within the range between zero and one. A topological phase transition has been predicted at $T=0$ in strongly correlated Fermi systems that is related to the emergence of a flat portion of the single-particle spectrum $\varepsilon(p)$ at the chemical potential. This transition is associated with the band flattening or swelling of the Fermi surface in momentum space, which is preceded by an increasing quasiparticle effective mass $m$ that diverges at the quantum critical point.

For an interacting fermion system with discrete spectrum, one expects a similar effect --- the appearance of different fillings of quantum levels at the chemical potential. Given the energies of two quantum levels intersect each other when varying an external parameter, these can be the same as the chemical potential over a range of parameter values, i.e., the levels can merge at the chemical potential over this range \cite{prl}. The level merging implies that there is attraction between the two partially-filled quantum levels. The merging interval is determined by a possibility of redistributing quasiparticles between the levels. It is clear that the effect of merging is in contrast to the simple crossing of quantum levels at some value of the external parameter.

Here, we show that the merging of the spin- and valley-split Landau levels at the chemical potential can be detected near the quantum critical point in a clean strongly-interacting two-dimensional (2D) electron system in (100) silicon. In this electron system subjected to perpendicular magnetic fields, each Landau level is split into four quantum levels due to the spin and valley splitting. It has been observed in experiment that the quantum oscillation minima at filling factor $\nu=4i+4$ (where $i=0,1,2,\ldots$) disappear below some electron density $n^*$ depending on $\nu$, while the minima at $\nu=4i+2$ persist down to appreciably lower densities \cite{krav}. Although this behavior is consistent with the sharp increase of the effective mass with decreasing electron density $n_s$ \cite{shashkin02}, the dependence of the density $n^*$ on filling factor (or magnetic field) turns out to be anomalously strong and lacks explanation. We find that the anomalous behavior of the density where the $\nu=4i+4$ oscillation minima vanish is described within the merging picture. This gives evidence for the level merging in a 2D electron system in silicon.

Imposition of the perpendicular magnetic field $B$ on a homogeneous 2D electron system is known to create two subsystems of Landau levels numbered $i$ and distinguished by $\pm$ projections of the electron spin on the field direction. For the time being, the valley degeneracy is neglected for the sake of simplicity. The energy levels $\varepsilon_i^\pm$ in each set are spaced by the cyclotron splitting $\hbar\omega_c=\hbar eB/mc$, and the two sets of the Landau levels are shifted with respect to each other by the spin splitting $\Delta_Z=g\mu_BB$, where $m$ and $g$ are the values of mass and Lande factor renormalized by electron interactions and $\mu_B=e\hbar/2m_ec$ is the Bohr magneton. Disregarding the anti-crossing effects, the Landau levels with opposite spin directions should intersect with changing electron density, as caused by the strong dependence of the effective mass on $n_s$, provided the $g$ factor depends weakly on $n_s$. In particular, at high electron densities the cyclotron splitting usually exceeds the spin splitting, whereas at low densities the opposite case $\hbar\omega_c<\Delta_Z$ should occur due to the sharply increasing mass. Below, we obtain conditions when the level crossing or level merging are possible.

Provided that the external magnetic field is fixed and weak, many quantum levels are occupied and the variation of the electron density in a quantum level is small compared to $n_s$. The variation of the energy $\varepsilon_\lambda$ is evaluated using the Landau relation
\begin{equation}
\delta\varepsilon_\lambda=\sum_\sigma\Gamma_{\lambda\sigma}\delta n_\sigma,\label{A}
\end{equation}
where $\Gamma_{\lambda\sigma}$ is the electron-electron interaction amplitude that is a phenomenological
ingredient of the Fermi liquid theory \cite{landau}. Selecting the magnetic field at which the difference between the neighboring Landau levels $\varepsilon_i^+$ and $\varepsilon_{i+1}^-$
\begin{equation}
D=\varepsilon_i^+-\varepsilon_{i+1}^-=\Delta_Z(n_s,B)-\hbar\omega_c(n_s,B)\label{dis}
\end{equation}
zeroes at the filling factor $\nu=n_shc/eB=2i+2$, we start from the higher density where both levels $(i+1)^-$ and $i^+$ are completely filled at $\nu=N=2i+3$, the difference $D(N)$ being negative. Removing the electrons from the level $(i+1)^-$ implies that the electron density decreases and $D$ increases. The above Landau relation reduces to the set
\begin{eqnarray}
\varepsilon_i^+-\varepsilon_i^+(N)&=&-(N-\nu)n_0\Gamma_{i,i+1}^{+-},\nonumber\\
\varepsilon_{i+1}^--\varepsilon_{i+1}^-(N)&=&-(N-\nu)n_0\Gamma_{i+1,i+1}^{--},\label{eps1}
\end{eqnarray}
where $\lambda,\sigma=i^+,(i+1)^-$ in the first equation, $\lambda,\sigma=(i+1)^-,(i+1)^-$ in the second one, and $n_0=eB/hc$ is the level degeneracy. Upon subtracting the second equation from the first, one has
\beq
\varepsilon_i^+-\varepsilon_{i+1}^-=D(N)+(N-\nu)n_0(\Gamma_{i+1,i+1}^{--}-\Gamma_{i,i+1}^{+-}).\label{dif1}
\eeq
The distance between the two levels vanishes as the level $(i+1)^-$ becomes empty, which corresponds to the relation
\beq
|D(N)|=n_0(\Gamma_{i+1,i+1}^{--}-\Gamma_{i,i+1}^{+-}).\label{ncr}
\eeq

Let us presume that at $\nu=2i+2$ the level crossing occurs, i.e., the level $i^+$ becomes empty and the level $(i+1)^-$ is completely filled. Then, the set (\ref{eps1}) should be replaced by
\begin{eqnarray}
\varepsilon_{i+1}^--\varepsilon_{i+1}^-(N)&=&-(N-\nu)n_0\Gamma_{i+1,i}^{-+}\nonumber\\
\varepsilon_i^+-\varepsilon_i^+(N)&=&-(N-\nu)n_0\Gamma_{i,i}^{++}\label{eps2}
\end{eqnarray}
to yield
\beq
\varepsilon_i^+-\varepsilon_{i+1}^-=D(N)+(N-\nu)n_0(\Gamma_{i+1,i}^{-+}-\Gamma_{i,i}^{++}).\label{dif2}
\eeq
Equations~(\ref{dif2}) and (\ref{dif1}) are compatible with one another, favoring the level crossing, under the condition $\Gamma(i)=(\Gamma_{i+1,i+1}^{--}-\Gamma_{i,i+1}^{+-})-(\Gamma_{i+1,i}^{-+}-\Gamma_{i,i}^{++})\leq0$.

In the opposite case
\beq
\Gamma(i)>0\label{ncm}
\eeq
the single-particle levels attract to each other and merge at the chemical potential $\mu$, as described by the merging equation $\varepsilon_{i+1}^-=\varepsilon_i^+=\mu$. Both levels exhibit partial occupation with fractions of empty states $0<f_i<1$ and $0<f_{i+1}<1$ that obey the normalization condition $f_i+f_{i+1}=f=N-\nu$. These fractions are determined from the set of equations
\begin{eqnarray}
\varepsilon_{i+1}^--\varepsilon_{i+1}^-(N)&=&-n_0(f_{i+1}\Gamma_{i+1,i+1}^{--}+f_i\Gamma_{i+1,i}^{-+})\nonumber\\
\varepsilon_i^+-\varepsilon_i^+(N)&=&-n_0(f_{i+1}\Gamma_{i,i+1}^{+-}+f_i\Gamma_{i,i}^{++})\label{epsc}
\end{eqnarray}
which yields
\begin{eqnarray}
\varepsilon_i^+-\varepsilon_{i+1}^-&=&D(N)+f_{i+1}n_0(\Gamma_{i+1,i+1}^{--}-\Gamma_{i,i+1}^{+-})\nonumber\\
&&+f_in_0(\Gamma_{i+1,i}^{-+}-\Gamma_{i,i}^{++}).\label{det}
\end{eqnarray}
Using the merging equation and the normalization condition, we find
\begin{eqnarray}
f_i&=&{(f-1)(\Gamma_{i+1,i+1}^{--}-\Gamma_{i,i+1}^{+-})\over\Gamma(i)}\nonumber\\
f_{i+1}&=&f-{(f-1)(\Gamma_{i+1,i+1}^{--}-\Gamma_{i,i+1}^{+-})\over\Gamma(i)}.\label{fc}
\end{eqnarray}
The merging starts when the empty states appear in the level $\varepsilon_i^+$ and ends when this level is completely emptied. This corresponds to the increase of the fraction of empty states $f$ in the range between $f=1$ (or $\nu=2i+2$) and $f=\min(1+\Gamma(i)/(\Gamma_{i+1,i+1}^{--}-\Gamma_{i,i+1}^{+-}),2)$. Outside the merging region, the conventional Landau level diagram is realized. Note that the gap between the neighboring Landau levels $\varepsilon_i^+$ and $\varepsilon_{i+1}^-$ proves to be invisible in transport and thermodynamic experiments. The upper boundary of the merging region $n_m(B)$ is written
\beq
\hbar\omega_c-\Delta_Z=0.\label{upper1}
\eeq

\begin{figure}
\scalebox{0.45}{\includegraphics{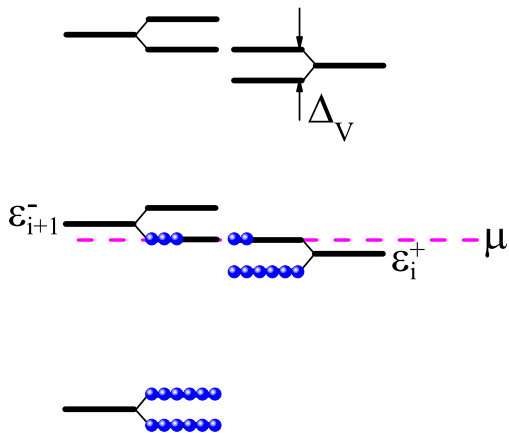}}
\caption{\label{fig1} Schematic diagram of merging of the spin- and valley-split Landau levels at the chemical potential. The occupied levels are indicated by dots. The fillings of the two quantum levels at the chemical potential vary with changing electron density.}
\end{figure}

Below, we compare the results of the calculations with the experimental data obtained in a strongly-interacting 2D electron system in (100) silicon. This electron system is characterized by the presence of two valleys in the spectrum so that each energy level $\varepsilon_i^\pm$ is split into two levels, as shown schematically in Fig.~\ref{fig1}. One can easily see that the valley splitting $\Delta_v$ promotes the merging of quantum levels. The bigger the valley splitting, the higher the electron density at which the levels $(i+1)^-$ and $i^+$ with different valley numbers should merge at the chemical potential at filling factor $\nu=4i+4$. The upper boundary of the merging region $n_m(B)$ is determined by the relation
\beq
\hbar\omega_c-\Delta_Z-\Delta_v=0\label{upper}
\eeq
that is different from Eq.~(\ref{upper1}) by the presence of the valley splitting. Since the electron density distributions corresponding to two valleys are spaced by distance $\alpha$ in the direction perpendicular to the Si-SiO$_2$ interface, the intervalley charge transfer creates an incremental electric field. In accordance with Eq.~(\ref{det}), we get $(\Gamma_{i+1,i+1}^{--}-\Gamma_{i,i+1}^{+-})=(\Gamma_{i,i}^{++}-\Gamma_{i+1,i}^{-+})=2\pi e^2\alpha/\kappa$ and $\Gamma(i)=4\pi e^2\alpha/\kappa$, where $\kappa$ is the dielectric constant. Although the distance $\alpha\sim0.4$~\AA\ is small compared to the thickness of the 2D electron system which is about 50~\AA\ at densities $n_s\approx1\times10^{11}$~cm$^{-2}$ \cite{ando}, the estimated interaction energy $n_0\Gamma(i)\sim0.02$~meV at $B\approx1$~T is comparable with the valley splitting $\Delta_v\approx0.06$~meV. The value of $\Delta_v$ is calculated using the known formula $\Delta_v\approx0.015(n_s+32n_{depl}/11)$~meV, where $n_{depl}\approx1\times10^{11}$~cm$^{-2}$ is the depletion layer density and the densities are in units of $10^{11}$~cm$^{-2}$ \cite{ando}. The strength of the merging effect being obviously determined by $\Gamma(i)$, the appreciable interaction energy should lead to a wide merging region at fixed filling factor $\nu=4i+4$. The lower boundary of the merging region of the neighboring Landau levels is given by the expression $\hbar\omega_c+n_0\Gamma(i)-\Delta_Z-\Delta_v=0$.

\begin{figure}
\scalebox{0.45}{\includegraphics{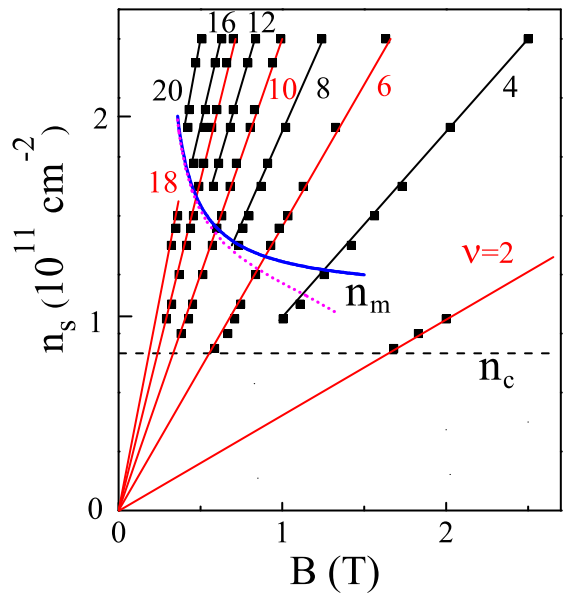}}
\caption{\label{fig2} Positions of the Shubnikov-de~Haas oscillation minima in the ($B,n_s$) plane (squares) and the expected positions of the cyclotron and spin minima calculated according to the formula $n_s=\nu eB/hc$ (solid lines). The position of the metal-insulator transition in $B=0$ is indicated. The calculated merging boundary $n_m(B)$ is shown by the solid blue line for $\beta=0$ and the dotted violet line for $\beta=1$ in Eq.~(\ref{m_line}). This makes sense at filling factor $\nu=4i+4$.}
\end{figure}

In the high-density limit, where the effect of electron-electron interactions is negligible, the effective mass and $g$ factor are equal to $m_b=0.19m_e$ and $g_0=2$ so that the cyclotron splitting significantly exceeds the spin splitting. At low electron densities, where the interaction effects are strong, the effective mass $m(n_s)$ is found to diverge as $m_b/m\simeq(n_s-n_c)/n_s$ at the quantum critical point close to the metal-insulator transition which occurs at $n_c\simeq 8\times 10^{10}$~cm$^{-2}$, while the $g$ factor stays close to $g_0$, being equal to $g\simeq 1.4g_0$ \cite{abrah,sar,shashrev,dolgop,shash2012}. The Landau level fan diagram for this electron system in perpendicular magnetic fields is represented in Fig.~\ref{fig2}. The quantum oscillation minima at filling factor $\nu=4i+4$ disappear below some electron density $n^*$ depending on $\nu$, while the minima at $\nu=4i+2$ persist down to appreciably lower densities \cite{krav}. Although this behavior is consistent with the sharp increase of the effective mass with decreasing $n_s$, the dependence of the density $n^*$ on filling factor (or $B$) turns out to be anomalously strong and lacks explanation. Particularly, this cannot be accounted by the impurity broadening of quantum levels in terms of $\omega_c\tau\sim1$ (where $\tau$ is the elastic scattering time) in which case the drop of mobility $e\tau/m$ at low electron densities is controlled by the increasing mass \cite{shashkin02}.

Using the above expressions for $m$, $g$, and $\Delta_v$, we determine from Eq.~(\ref{upper}) the expected upper boundary of the merging region $n_m(B)$, shown by the solid blue line in Fig.~\ref{fig2}. The calculated boundary is in agreement with the experimental density $n^*(B)$ at which the oscillation minima at $\nu=4i+4$ vanish. This fact gives evidence for the level merging in a 2D electron system in silicon.

We now discuss the possibility that the description of the high-field data $n^*(B)$ improves within the merging picture if one takes account of nonlinear (cubic) corrections to the spectrum $\varepsilon(p)$ near the Fermi surface that lead naturally to a decrease of the effective mass with magnetic field. The cubic corrections should be important near the quantum critical point since the linear term is strongly suppressed, and the spectrum takes the form \cite{prb2008}
\beq
\varepsilon(p)-\mu={p_F(p-p_F)\over m}+\beta{(p-p_F)^3\over 3m_bp_F},\label{cubic}
\eeq
where $p_F=\hbar(\pi n_s)^{1/2}$ is the Fermi momentum and $\beta>0$ is a coefficient. The cubic correction corresponds to an additional term in the single-particle Hamiltonian in magnetic fields. According to textbook rules, this term is written
\beq
{\cal H}_{add}=\beta{\left[({\bf p}{-}e{\bf A}/c)^2{-}p^2_F\right]^3\over 24m_bp_F^4},\label{cubicH}
\eeq
where ${\bf A}$ is the vector potential. The resulting correction to the spectrum leads to a modification of Eq.~(\ref{upper}) that describes the upper boundary of the merging region. The corrected equation reads
\beq
\hbar\omega_c-\Delta_Z-\Delta_v=-\beta{(eB)^3\over 3m_b\hbar c^3(\pi n_m)^2}.\label{m_line}
\eeq
Apparently, the right hand side of Eq.~(\ref{m_line}) can be important at high magnetic fields. Assuming that the coefficient $\beta=1$, we estimate the influence of the correction and determine the corrected dependence $n_m(B)$, shown by the dotted violet line in Fig.~\ref{fig2}. One concludes that the experimental data for the density $n^*(B)$ at which the oscillation minima at $\nu=4i+4$ disappear can be even better described within the concept of merging by taking into account cubic corrections to the spectrum at the Fermi surface near the quantum critical point.

In summary, we have shown that the merging of the spin- and valley-split Landau levels at the chemical potential in a clean strongly-interacting 2D electron system in silicon is confirmed by available experimental data.

This work was supported by RFBR 12-02-00272, 13-02-00095, and 14-02-00044, RAS, the Russian Ministry of Sciences, NS-932.2014.2, RCSF 14-12-00450, the U.S. DOE, Division of Chemical Sciences, the Office of Basic Energy Sciences, the Office of Energy Research, AFOSR, and the McDonnell Center for the Space Sciences.

\end{document}